\documentclass[conference]{IEEEtran}
\IEEEoverridecommandlockouts

\usepackage{cite}
\usepackage{amsmath,amssymb,amsfonts}
\usepackage{graphicx}
\usepackage{textcomp}
\usepackage{xcolor}

\usepackage{algorithm}
\usepackage{algpseudocode}
\usepackage{subcaption}
\usepackage{float}
\usepackage[normalem]{ulem}

\usepackage{subfiles}
\usepackage{pdfpages}

\usepackage{setspace}
\usepackage{pifont}
\usepackage{color}
\usepackage{booktabs}
\usepackage{enumitem}
\usepackage{hyperref}

\usepackage[framemethod=tikz]{mdframed}
\def\BibTeX{{\rm B\kern-.05em{\sc i\kern-.025em b}\kern-.08em
    T\kern-.1667em\lower.7ex\hbox{E}\kern-.125emX}}
    
\begin{document}

\newcommand{\ka}[1]{{\color{red}{} #1}}
\newcommand{\kr}[1]{{\color{blue}{} #1}}
\newcommand{\jz}[1]{{\color{orange}{} #1}}
\newcommand{\fc}[1]{{\color{magenta}{} #1}}

\newcommand{\method}{SAIL}

\title{\method: Self-Adaptive Index Learning with Hybrid Models (Extended Version)\\
}

\author{\IEEEauthorblockN{Kasun Amarasinghe}
\IEEEauthorblockA{
\textit{The University of Melbourne}\\
Melbourne, Australia \\
kasun.amarasinghe@student.unimelb.edu.au}
\and
\IEEEauthorblockN{Farhana Choudhury}
\IEEEauthorblockA{
\textit{The University of Melbourne}\\
Melbourne, Australia \\
farhana.choudhury@unimelb.edu.au}
\and
\IEEEauthorblockN{Jianzhong Qi}
\IEEEauthorblockA{
\textit{The University of Melbourne}\\
Melbourne, Australia \\
jianzhong.qi@unimelb.edu.au}
}

\maketitle

\begin{abstract}
Learned index structures have demonstrated a significant improvement in query performance over traditional index structures.
The majority of learned indexes use a uniform node structure throughout their index structures. A few studies show that combining multiple types of node structures can lead to indexes that better adapt to local key-set distributions, hence yielding query efficiency gains. 
Following this line of work, we propose a learned index structure named \method\ that supports a wide range of different candidate node structures for both leaf and inner nodes, selected via cost models based on data driven approximations. We design an efficient construction algorithm that recursively partitions the key-set to be fitted by the candidate node types. Experiments demonstrate that \method\ achieves 2x improvement over existing learned indexes in query time, with similar memory consumption and index build times. 

\end{abstract}

\section{Introduction}
Modern big data analytics platforms and cloud data systems rely heavily on index structures to deliver high throughput query performance over large datasets~\cite{dynamoDB,RocksDB}. Recent research has shown the benefits of using learned indexes instead of traditional index structures~\cite{kraska2018case}. As data volumes continue to grow, learned indexes have shown better adaptability to key-set distributions to deliver superior query performance and as such are an active and growing area within the database community.

The main idea of a learned index is to create a mapping function of $f(\cdot)$ for a querying key to the storage location, such that query processing is done with function invocation instead of traversals over an index structure.  For the creation of such mapping functions, a.k.a. \emph{indexing functions}, a storage ordering is required. This is commonly done by ascending order of the keys, hence the indexing functions are typically the rank $rank(k_i)$ of the data, $rank(k_i)\approx f(k_i)$, making them an approximation of the \emph{cumulative distribution function} (CDF) of the data keys. Machine learning approaches are generally used to learn the functions.

\begin{figure}[t]
  \centering
  \begin{subfigure}[b]{0.48\linewidth}
    \includegraphics[width=\linewidth]{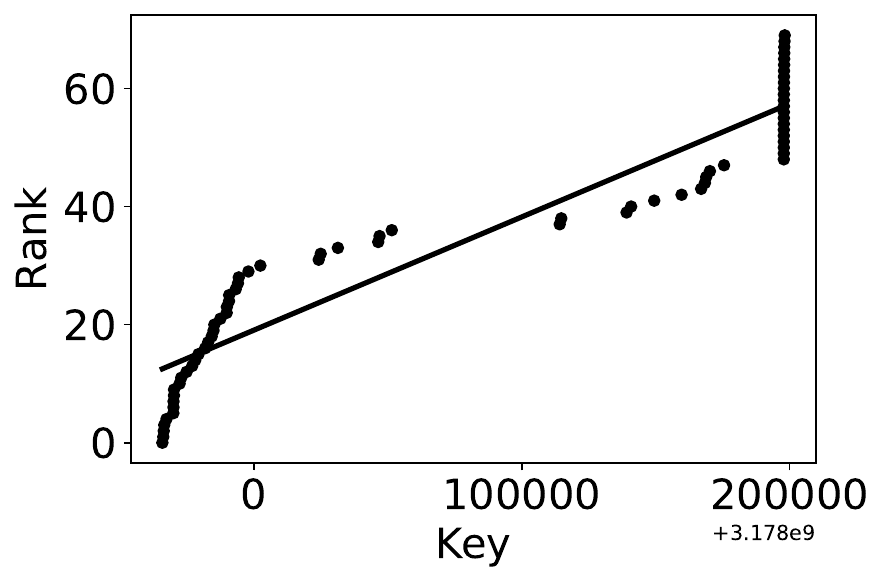}
    \caption{Single indexing model}
    \label{fig:partition_keys_a}
  \end{subfigure}
  \hfill
  \begin{subfigure}[b]{0.48\linewidth}
    \includegraphics[width=\linewidth]{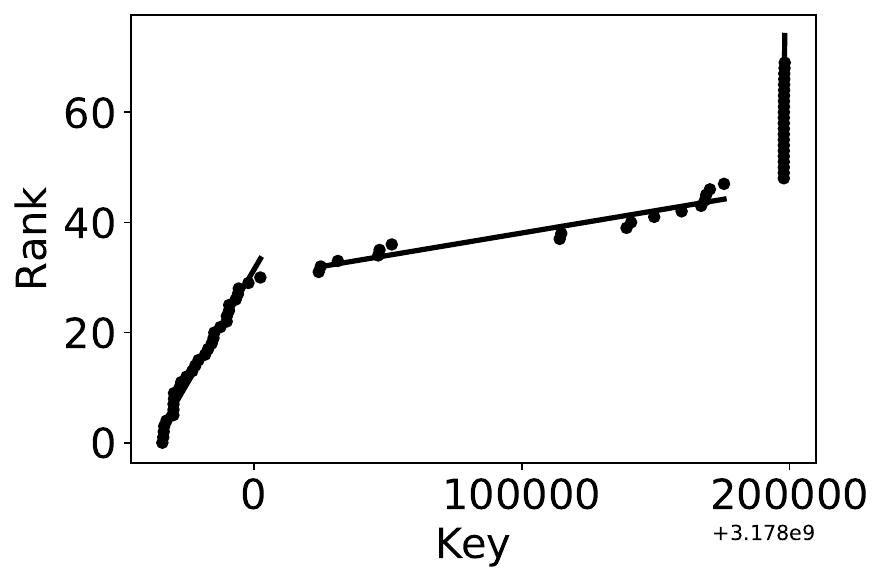}
    \caption{Multiple indexing models}
    \label{fig:partition_keys_b}
  \end{subfigure}
  \caption{Partitioning key-set for indexing function fitting.}
  \label{fig:partition_keys}
\end{figure}

A rich set of different learned indexes have been proposed~\cite{amarasinghe2025learned, ding2020alex, ferragina2020pgm,wu2021lipp, zhang2021colin,nathan2020flood, li2020lisa, ding2020tsunami, ding2022hybrid_error, pai2023workloadz, kraska2018case, sheng2023wisk, li2021finedex, tang2020xindex, ge2023sali, li2025vega, guo2024chameleon}. Beyond efficient indexing of key sets, these studies have also considered insertion capabilities and better approximation of the CDFs when the key-set distributions are difficult to learn. While studies have examined the effectiveness of these structures over different datasets~\cite{kipf2019sosd,wongkham2022updatable, sun2023learned}, \emph{the strength of different learned indexes over different subsections of the same dataset has been largely overlooked.}

Fig.~\ref{fig:partition_keys} presents an example from the Genome dataset (detailed in Section~\ref{sec:experiments}), where the key space depicts three distinct segments. Using a single linear regression-based indexing function across all three segments (cf.~Fig.~\ref{fig:partition_keys_a}) results in a \emph{maximum error} of 12, i.e., the largest difference between the actual and the predicted position of any keys. This would translate to poor query performance if used for indexing. Fitting each segment individually with regression-based functions reduces the maximum error to 3, 2, and 6 respectively (cf.~Fig.~\ref{fig:partition_keys_b}), which is the idea of existing works that use piecewise learned indexing functions~\cite{ferragina2020pgm,galakatos2019fiting}. 

We observe that the maximum error could be further reduced by utilizing different types of indexing functions for the different segments. For example, hashing, which is used by the state-of-the-art (SOTA) approach \cite{guo2024chameleon}, would suit the middle segment nicely, while it would not suit the segment on the right as the keys will be mapped to the same position and result in collisions. 

Motivated by this observation, we aim to design an index structure that utilizes existing or emerging, traditional and learned indexes for different partitions of a key-set to optimize query efficiency and streamline big data analytics. 
Rather than introducing another new kind of indexing function, we focus on designing an index structure that \emph{adaptively partitions the input dataset into sections and choosing a query efficient indexing function for each section}. Specifically, we formulate the problem as the selection of the best partitions and node type (i.e., the type of indexing functions or traditional index nodes) for each partition for efficient query performance. 

A major challenge is the combinatorially high search space formed by the many partitions and the choice of node/function types. It requires an effective mechanism to quickly estimate the query performance of each node type for comparing the choice of different index node/function types and the decision to whether further partition a sub-key-set or directly store it as a leaf node at each step. 

To address this challenge, we introduce cost models that model query costs based on core operations needed to perform index querying such as integer comparisons and arithmetic operations, calibrated to specific hardware configurations. These cost models guide a recursive top-down partitioning of the key-set to allocate the appropriate node structure at each step. To reduce the number of possible partitions, we propose a new type of cache-efficient node that includes a ``sub-partition'' called buckets, grouping key-set regions with complex local distributions that cannot be well approximated by a single learned model. We call our resulting index structure \textbf{\method} for its \underline{s}elf-\underline{a}daptive \underline{i}ndex \underline{l}earning capability.

A benefit of our design is that our solution supports range queries by preserving the keys in sorted order, which is particularly useful for big data analytical queries~\cite{aly2016kangaroo,li2014fastrange}, while it also achieves SOTA point query performance. 

To summarize, this paper makes the following contributions: 

\begin{enumerate}[leftmargin=*]
\item We propose a novel index construction technique named \method\ 
that adaptively partitions an input dataset and chooses the most appropriate indexing function/node structure for each partition for query time optimization. 

\item We propose a new inner and leaf node hybrid that indexes complex local distributions with fewer partitions. 

\item We propose cost models to efficiently evaluate the query performance of different node types, which are generic in design based on core operations to accommodate different existing and possible emerging indexes.

\item We conduct extensive experiments and compare our index structure with SOTA learned index structures on four real datasets and two synthetic datasets. The experimental results show  a 2x performance gain against SOTA index structures for point queries. For range queries, the performance is on par with existing best performing indexes. Among the learned indexes evaluated, there is \emph{no single existing index with such consistent performance for both point and range queries}, while such strong query performance is achieved together with competitive build times and memory consumptions. 
\end{enumerate}

\section{Related Work}

\textbf{Learned indexes.} 
Learned index structures are widely studied in the database community recently~\cite{wu2021lipp,guo2024chameleon,zhang2026hire}. These indexes view their structures as functions that map keys to their respective storage locations. These indexing functions are typically created using machine learning algorithms based on key-sets that are sorted in ascending order, thereby effectively learning the CDFs of the keys. Different types of indexing functions have been developed, each suiting different data distributions~\cite{ding2020alex,wu2021lipp,guo2024chameleon}. In real datasets, the data distribution is often not uniform, with local distributional variances that challenge the use of a single type of indexing function throughout an entire index structure. Inspired by this observation, we propose to utilize different indexing functions within a hybrid index structure to fit varying local distributions of a dataset. 

\textbf{Learned indexes for complex data distributions.} 
To index key-sets with complex distributions, a common approach is to partition and use a different indexing function (of the same type) for each partition, which is then organized in a hierarchical structure~\cite{ferragina2020pgm,galakatos2019fiting}. Another approach is to use more complex indexing functions, such as splines~\cite{kipf2020radixspline} and higher polynomial regressions~\cite{li2021polyfit}, to approximate more complex dataset distributions. Both approaches come with costs. Using more complex indexing functions tend to increase the training and inference (i.e., query) times, while partitioning data results in deeper index structures that cause higher traversal times. 
For example, VEGA~\cite{li2025vega} divides the key-set into groups and learns the indexing model based on one key from each group, instead of the full dataset. It is easier to build models that fit fewer data points, thus reducing the number of models and the height of the index. At query time, the models are used to identify the group to which the queried key belongs. Then, Single Instruction, Multiple Data (SIMD) is used to search within the group. This was designed to be a read-only index.  
Chameleon~\cite{guo2024chameleon} uses reinforcement learning (RL) to identify locally skewed regions and attempts to divide the key-set for such regions. 
It uses an Actor-Critic model to generate a structure to set the fan-out of each index node. A Genetic algorithm is used as the Actor, while a trained Deep-Q network is used as the Critic. A second RL agent performs a fine-grained assessment of the leaf nodes to evaluate whether further divisions are required. All leaf nodes are created with an error-bounded hashing function. While Chameleon has high point query efficiency, the use of error-bounded hashing hinders the range query efficiency. 

\textbf{Hybrid learned indexes.} 
UPLIF~\cite{heidari2024uplif} uses RL to optimize the index structure by choosing between\ two traditional structures, Red-Black-Tree and B+-tree. Unlike our proposal, UPLIF does not consider data partitioning along with structure selection. CARMI~\cite{zhang2021carmi} uses entropy based cost models and dynamic programming to select between different node types. This limits the possible node types because the entropy model requires all candidate nodes to preserve data order. CARMI sets the node fan-outs by iterating through several values, which makes it computationally difficult to find the optimal fan-out without significant index build times. Shape-shift~\cite{wang2025shapeshifter} utilizes two different structures, B+-tree and a learned structure. It uses a cost calculation to determine the structure to be selected. 
Another study attempts to combine the benefits of traditional and learned indexes by creating a hybrid index structure~\cite{zhang2026hire}. It also uses a cost model-based approach to determine whether to use linear regression model-based leaf nodes or B+-tree style leaf nodes. These approaches limit their nodes to two types.

A self-designing index, GENE~\cite{dittrich2021gene}, uses genetic algorithms to randomly mutate and generate the indexes, while SELIX~\cite{han2026selix} trains an RL agent and uses a unified index template to obtain the node layouts. Both approaches rely on computationally expensive searches (RL and genetic algorithms) to generate the indexes, limiting their adaptability to new node types.

Our proposed approach differs from these studies as we aim to develop an adaptive, extendable structure that could add multiple different leaf or inner node types as desired. The literature has typically used a binary choice of node types or requires computationally expensive retraining, while ours allows for more node types, hence opening up opportunities for query efficiency optimization. 

\begin{figure*}[t]
  \centering
    \includegraphics[width=\linewidth]{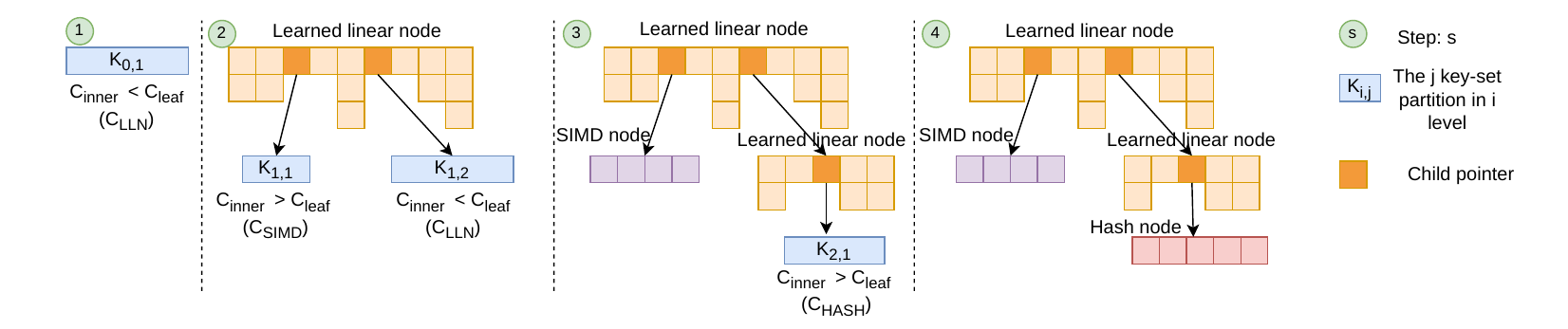}
  \caption{Our \method\ index structure. (1)~Costs are calculated for the key-set $K_{0,1}$ as described in Section~\ref{sec:cost_models}. It creates a learned linear node which has the smallest cost. (2)~This results in two further key-set partitions, $K_{1,1}$ and $K_{1,2}$. The recursive operation is run on these two partitions. (3)~The $K_{1,1}$ key-set partition is stored in an SIMD search leaf node, while $K_{1,2}$ requires further partitioning. (4)~The $K_{2,1}$ key-set partition creates a hashing-based leaf node, completing the index structure.}
 \label{fig:ideal_structure}
\end{figure*}

\section{Preliminaries} \label{sec:preliminaries}

\textbf{Problem statement.} 
\sloppy
Given a sorted one-dimensional key-set $K$ of size $n$ with key-value pairs ${(k_1,v_1), (k_2,v_2), \dots , (k_n,v_n)}$, our aim is to construct an in-memory index structure using a variety of node types taken from a set of available node types $M$. We propose a recursive partitioning and a node type assignment strategy such that the resulting index structure is optimized for query performance. We focus on point and range queries as these are the fundamental query types.   

\textbf{Challenges.} 
To create the proposed solution, there are several challenges to be addressed: (1)~\uline{Combinatorially high search space.}
The partitioning can be done at any location for any number of partitions. This leads to a exponentially high number of possible partitioning candidates.  (2)~\uline{Node type and partitioning dependency.}
The decision to further partition a sub-key-set or to directly store it as a leaf node cannot be determined by the number of keys alone. Unnecessary partitions could lead to a deep index structure and longer traversal times during query processing. Similarly, the use of an incorrect node type would cause potentially high query times. As such, the partitioning choice and the node type assignments are correlated, which further complicates the problem. 
(3)~\uline{Cache efficiency.} 
When a partition exceeds the cache capacity, it degrades the query performance as cache misses occur at query time. Therefore, the design should further consider cache efficiency to deliver better performance. (4)~\uline{Range query and data insertion support.}
Range queries and data insertions are common for big data analytics. A well constructed index should support not only point queries (as done in many learned index studies~\cite{wu2021lipp,ding2020alex,ferragina2020pgm}) but also range queries and data insertions efficiently.

\section{\method} \label{sec:sys_overview}
To overcome these challenges, our \method\ index, as plotted in Fig.~\ref{fig:ideal_structure}, is constructed with a top-down recursive procedure. It selects the most appropriate node type for specific key partitions from a set of candidate structures via a cost model. Rather than using a single node type for the entire index structure, our structure assigns different node types per key-set partition by examining the properties of the local keys. We detail the node types, their cost models, \method\ construction, query, and update handling procedures in this section.

\subsection{Candidate Node Types}
\label{sec:node_types}
\method\ uses the following four node types: binary search leaf node, SIMD search leaf node, hashing-based node, and \emph{a new ``learned linear inner node''} that we propose to reduce the number of partitions (detailed in Section~\ref{sec:lln_node}). The learned linear nodes and hashing-based nodes can act as either inner nodes or leaf nodes.  We have chosen these four node types as they consist of both traditional methods such as binary search and hashing, as well as the latest indexing mechanisms such as SIMD search and learned indexing functions to address the needs of different key-set distributions. Our index structure is generalizable to other node types by the inclusion of relevant cost models.

\subsubsection{Learned Linear Inner Node} \label{sec:lln_node}
Learned linear nodes can act as inner nodes that direct to child nodes at next levels and leaf nodes that predict data storage locations, simultaneously. This is achieved by storing keys within an inner node itself. The node design was inspired by the LIPP index~\cite{wu2021lipp}, which has no searching component. 

Similar to the LIPP index, we collect the conflicting keys predicted to the same location by a linear model and create a new node at the next level. The process is done until there are no more conflicting keys. For this purpose, a linear model to minimize the maximum number of prediction collisions is needed instead of minimizing the mean squared error.  This is beneficial as it allows more keys to be placed and retrieved from the node itself.
Since the linear model used is monotone, the keys that collide are consecutive. This allows the keys to be used for child node construction at the next level and support efficient range query processing.  

For our purpose, directly using the original LIPP node was not viable due to several limitations. The original node design only allows for a single key to be placed at one location, limiting its efficiency, as even just two keys colliding would require a new child node to be created, leading to deeper index structures. To mitigate this issue, 
LIPP sets a higher node capacity to allow for more locations without collisions. Furthermore, its use of separated metadata arrays add cache misses during query processing. Our new learned linear node overcomes these limitations with the following changes. 

\begin{itemize}[leftmargin=*]
    \item We include an additional storage of size $B$ (like separate chaining to handle collisions in hashing), called \emph{buckets}, within the inner nodes to accommodate the colliding keys at each location (only when there is at least one collision), helping our \method\ index to keep a lower structure. These act as a sub-partition structure for the inner nodes to handle small partitions within the node itself and deferring child node creation only for partitions that overflow the buckets.
    \item We design a cache-efficient layout by aligning each location, a.k.a. a slot, to a 64 byte cache line, that stores all the necessary information including the key-value pairs, number of key-value pairs in the location, and the overflow child pointer. If the search key is in a bucket, the lookup operation can be performed with one cache line load.
\end{itemize}

\subsubsection{Binary Search Leaf Node} \label{sec:binary_node}
A binary search node stores two sorted arrays as a leaf node. It holds the keys and values in two parallel sorted arrays. This serves as the fallback node type when the other types are ill fitted. At query time, a binary search is run to find the key-value pair that matches the query key for such a node.

\subsubsection{SIMD Search Leaf Node} \label{sec:SIMD_node}

Similarly, an SIMD search leaf node stores sorted arrays as a leaf node as well. This leaf node performs point queries by vectorized equality comparison. AVX-512 enables the leaf to compare eight 64-bit keys in a single instruction. SIMD search nodes outperform binary search nodes for a small (up to 16) number of keys~\cite{li2025vega}. 

\subsubsection{Hashing-based Leaf Node} \label{sec:hash_node}
Hashing-based search nodes place keys at the location determined by a hash function. We use  MurmurHash3 64-bit finalizer as it consists of simple operations for execution efficiency~\cite{sabek2022canhash}. If the location already holds a key, we use linear probing in both directions to find the closest available location to index the key. The maximum distance needed for linear probing is recorded during index construction to bound query search process. Such nodes cannot maintain the keys in sorted order, which could negatively impact range query efficiency. As such, only small key segments (with less than a threshold number of keys) are considered for this node type for leaf nodes. 

\subsubsection{Hashing-based Inner Node}
To show the generalizability of our structure, we also use inner nodes  with hashing (MurmurHash3 64-bit finalizer). Similar to the the learned linear inner nodes, hashing-based inner nodes use sub-partition support via buckets of size $B$, and the nodes are implemented in a cache-efficient manner. During our evaluations in Section~\ref{sec:experiments}, we demonstrate that the inclusion of this inner node type along with the learned linear inner nodes significantly improves the point query performance, at the cost of range query performance.

The most accurate way of assigning the correct node types would be to construct an index with each of the node types available for the key-set over any given partition and to measure the query performance empirically. However, doing so would be computationally infeasible. To overcome this issue, we develop cost models using data-driven information, the structural information of the node types as well as hardware settings to estimate the query costs. The construction decision then becomes a selection among the different types of nodes using these cost models. 

\subsection{Cost Models for Node Types}
\label{sec:cost_models}

Our cost model estimates query time as a weighted sum of core cost factors calibrated on target hardware:
\begin{itemize}[leftmargin=*,label={}]
    \item $C_\text{compare}$: cost for an integer comparison operation. It is the baseline cost used to measure other cost factors; 
    \item $C_\text{mult\_add}$: relative cost compared to $C_\text{compare}$ for multiplication and addition operations. This cost is needed for linear model evaluation;
    \item $C_\text{hash}$: relative cost compared to $C_\text{compare}$ for hashing;
    \item $C_\text{cache}$: relative cost compared to $C_\text{compare}$ for a cache miss;
    \item $C_\text{rate}$: cache miss rate as a percentage;
    \item $C_\text{setup}$: relative cost compared to $C_\text{compare}$ for the setup cost for SIMD search.
\end{itemize}

The cost model uses average-case estimates, which align with throughput oriented workloads typical of big data analytics.

\subsubsection{Learned Linear Inner Node Cost}
A learned linear node has two possible paths for query processing: (1) the queried key is stored in the bucket of the predicted location, or (2) the queried key is in a child node at the location when the bucket overflows. Our cost model is created to reflect both scenarios. These two paths are weighted by the fraction of keys in each. 

\textbf{Bucket storage path cost.} In this case, the queried key is used to predict the location via a linear model. The linear model evaluation takes $C_\text{mult\_add}$ cost. After finding the location via the prediction, the cache-efficient layout is loaded into memory at the cost of $C_\text{cache}$. Finally, scanning the bucket would cost $(B/2) \cdot C_\text{compare}$ on average, where $B$ is the size of the bucket storage. Overall, the cost is: 
\begin{equation}\label{eq:inline_cost}
Cost_\text{LLN\_bucket}(n) = C_\text{mult\_add} + C_\text{cache} + (B/2) \cdot C_\text{compare}
\end{equation}

\textbf{Child node path cost.} In this case, the queried key needs to also check the bucket storage. Therefore, the $Cost_\text{LLN\_bucket}(n)$ cost is included. Then, a further $C_\text{cache}$ cost would incur to load the child node structure into cache. Finally, we need to approximate the search cost of the child node. It would be computationally expensive to calculate the costs for each child node type, which is also unnecessary as our system recursively checks each of the child nodes later on. The constant 4 used here approximates four child nodes created by the bucket overflow as the majority of the child nodes using learned linear nodes have small numbers of keys in them. Overall, the cost of accessing a leaf linear node is: 

\begin{equation}\label{eq:child_path_cost}
Cost_\text{LLN\_child}(n) = Cost_\text{LLN\_bucket}(n) + C_\text{cache} + 4 \cdot C_\text{compare}
\end{equation}

\textbf{Learned linear inner node total cost.} The cost of a learned linear model is calculated by multiplying the bucket storage path cost and the child node path cost by its respective number of keys in each path as a fraction. These two fractions are obtained by fitting the linear model on the key-set and calculating how many fall within the bucket storage or overflow to child nodes, empirically. This is done to obtain an accurate estimate for the cost rather than using a theoretical assumption for the learned linear node type. 
Therefore, our cost model captures how partitioning would actually impact the index structure by using this observed measurement of the key-set. 
\begin{equation}\label{eq:llr_cost}
Cost_\text{LLN}(n) = f_\text{bucket} \cdot Cost_\text{LLN\_bucket}(n) + f_\text{child} \cdot Cost_\text{LLN\_child}(n) 
\end{equation}
where $f_\text{bucket}$ and $f_\text{child}$ are the fractions for the bucket storage keys and the child nodes, respectively.

\subsubsection{Hashing-based Inner Node Cost}
The same principles used for learned linear inner node's cost modeling applies to the hashing-based nodes. There are two possible paths where a key could be indexed, either in the bucket storage, or as a child node. The bucket storage path's cost calculation remains the same as that for the learned linear inner nodes, except for the model calling costs. For hashing-based nodes, the model calling cost is denoted as $C_\text{hash}$. The overall cost becomes: 
\begin{equation}\label{eq:inline_cost_hash}
Cost_\text{HASH\_bucket}(n) = C_\text{hash} + C_\text{cache} + (B/2) \cdot C_\text{compare}
\end{equation}

For cases where the queried key resides in a child node, the cost calculation is again similar to that for the learned liner nodes, except that now $Cost_\text{HASH\_bucket}(n)$ is used.  
\begin{equation}\label{eq:child_path_cost2}
Cost_\text{HASH\_child}(n) = Cost_\text{HASH\_bucket}(n) + C_\text{cache} + 4 \cdot C_\text{compare}
\end{equation}

The total cost for hashing-based inner nodes is the sum of these two costs multiplied by the fraction of keys in each path. The fraction of keys are obtained as described before.

\begin{equation}\label{eq:llr_costh}
\begin{split}
Cost_\text{HASH}(n) = f_\text{bucket} \cdot Cost_\text{HASH\_bucket}(n) \\+ f_\text{child} \cdot Cost_\text{HASH\_child}(n) 
\end{split}
\end{equation}
Following a similar approach, \method\ can  easily incorporate new node types (indexing functions) by creating their respective cost models.

\subsubsection{Binary Search Leaf Node Cost}
For a key-set of size $n$, binary search takes approximately $ \log_2(n+1)$ comparisons to  search for a key. We also include additional costs due to cache misses in the overall cost:  
\begin{equation}\label{eq:bs_cost}
Cost_\text{BS}(n) = \log_2(n+1) \cdot (C_\text{compare} + C_\text{rate} \cdot C_\text{cache})
\end{equation}

\subsubsection{SIMD Search Leaf Node Cost}
SIMD search nodes require the queried key to be broadcast to the vector register, which costs $C_\text{setup}$. Each iteration of the scans performs load and compare operations, the costs of which are given as $2 \cdot C_\text{compare}$. For a key-set of size $n$, the number of iterations is $ \lceil n/w \rceil$, where $w$ is the width of the SIMD vector. Overall, the query cost for an SIMD node is: 
\begin{equation}\label{eq:simd_cost}
Cost_\text{SIMD}(n) = C_\text{setup} + \lceil n/w \rceil \cdot 2 \cdot C_\text{compare}\end{equation}

\subsubsection{Hashing-based Leaf Node Cost}
A hashing-based leaf node first calculates the hash of the queried key at the cost of $C_\text{hash}$. Afterwards, a bidirectional bounded linear probe is performed within $\epsilon$ locations in both directions ($\epsilon = \log_2(n) + 2$). The cost for this probing is estimated as $(2 \cdot \epsilon + 1) \cdot C_\text{compare}$. Overall, the query cost of a hashing-based leaf node is: 
\begin{equation}\label{eq:hash_cost}
Cost_\text{HASH}(n) = C_\text{hash} + (2 \cdot \epsilon + 1) \cdot C_\text{compare}
\end{equation}

\subsection{Index Construction}
\begin{algorithm}
\begin{small}
\caption{Build($K$)}\label{alg:const}
\begin{algorithmic}[1]
\Require Key-set: $K$,
\Ensure Root node
\ForAll{$\text{inner\_node} \in \text{inner\_nodes}$}
\State Get $model$ = $Inner\_model(K,\text{inner\_node})$
\State Get $location\_keys$ = $predict(model)$
\State $f_\text{bucket}, f_\text{child}$ = compute keys fractions from 
$location\_keys$
\State Calculate $Cost_\text{inner\_node}$
\EndFor
\State $Cost_\text{inner}$ = $minimum(Cost_\text{inner\_nodes})$
\State Calculate leaf costs, $Cost_\text{SIMD}, Cost_\text{HASH} \text{ and } Cost_\text{BS}$
\State $Cost_\text{leaf}$ = $minimum(Cost_\text{SIMD}, Cost_\text{HASH}, Cost_\text{BS})$ 
\If{$Cost_\text{leaf}$ < $Cost_\text{inner}$}
  \State \Return $Build\_best\_leaf(K)$
\EndIf
\State $inner\_node$ = $Build\_inner\_node(K)$
 \ForAll{$loc \in $\text{ locations of inner\_node}}
    \If{${location\_keys}(loc).size \leq B$}
        \State Place ${location\_keys}(loc)$ in bucket storage
    \Else
        \State $loc.child\_pointer$ = $Build({location\_keys}(loc))$
    \EndIf
\EndFor
\State  \Return $inner\_node$
\end{algorithmic}
\end{small}
\end{algorithm}

As summarized in Algorithm~\ref{alg:const}, our \method\ index structure is constructed recursively in a top-down manner. At each node, our build function is executed to determine whether to further split or assign a leaf node type. Our structure terminates its construction locally when the key-set in that region is properly fitted. The choice to whether partition further is guided by the cost models discussed in Section~\ref{sec:cost_models}. 

First, for each of the inner node types in our candidate set of nodes, we fit the corresponding model.
The fitted model is used on these keys to obtain the actual locations where the keys are predicted as a location occupancy distribution known as $location\_keys$. This measurement obtained by the data itself enables a more accurate cost model computation than a theoretical estimate alone, as it captures how a potential partitioning would impact a specific key-set. The fraction of keys that overflows the bucket storage of size $B$ to create child nodes and those that are placed within the inner node's buckets themselves can easily be calculated from this distribution. Afterwards, as shown at Line 5, the cost for the inner node is calculated via the cost models described in Section~\ref{sec:cost_models}. This process is performed for all inner node types. The node with the smallest cost is chosen as the potential inner node. 

Then, the leaf node costs of different node types are calculated as described in Section~\ref{sec:cost_models}. The minimum is chosen as the cost for the leaf node (Lines 8-9). The decision to partition or not is made by comparing these costs (Lines 10-12). If the cost of the leaf node is less than that of the inner node, we do not  partition further but create a leaf node. 

Line 13 and below describe the partitioning and recursive aspects of our index building. After creating an inner node using the key-set, each location of this node is traversed. With the aid of $location\_keys$, the keys are placed in the bucket storage if the size is less than the storage size $B$, as shown at Line 16. Otherwise, as given at Line 18, the index building function is called recursively for the keys in that overflow location. A pointer to this child node is saved at the location. 
This recursion creates shallower index structures when the key-set is more uniform, while allowing further partitions when it is not. By utilizing a candidate set of node types to find the appropriate node type for specific segments of the key-set, our \method\ index structure adapts to local key-set distributions.

An example of the index building process is given in Fig.~\ref{fig:ideal_structure}. Costs are calculated for the full key-set ($K_{0,1}$) and a learned linear inner node is chosen as the best node type. It stores several keys within the node structure itself, and also creates two key-set partitions, $K_{1,1}$ and $K_{1,2}$. The first partition creates an SIMD search leaf node, while the second requires further partitioning. As such, another inner node is created for a key-set partition $K_{2,1}$. A hashing-based leaf node is created for  this partition, and the index construction completes.

\subsection{Query and Update Processing}
\textbf{Point queries.}
Point queries start from the root node.  Inner nodes use their indexing functions to predict the search key's location. If the location has not overflowed to a child node, the key-value pair is retrieved by scanning the storage bucket. Otherwise, the child node is called recursively. Leaf nodes locate the key by using their respective mechanism, linear model, binary search, SIMD search, or hashing with a bounded search (cf.~Section~\ref{sec:node_types}). Since the linear model is used only for routing and each prediction is followed by a bucket scan or a child lookup, no key is missed.

\textbf{Range queries.}
Range queries follow the same top-down traversal as point queries. A learned linear inner node predicts both the start and end key locations. Since the model is monotone, the required keys fall entirely within this range. If the keys are placed in storage buckets, the keys within the range are returned. Otherwise, the respective child node is called. For binary search and SIMD nodes, the starting key is located and the array is traversed until a key exceeding the end key is reached. Hashing-based leaf nodes do not preserve order, as such the complete node is traversed.

\textbf{Update handling.}
An insertion performs a point query to obtain the location of the insertion key. If the key already exists, its value is updated. For inner nodes, if the bucket at the predicted location has remaining capacity, the key-value pair is stored directly. Otherwise, the keys at the location are converted into a new child node. Leaf nodes use their respective insertion strategies, binary search and SIMD nodes shift the keys to maintain sorted order, while hashing-based leaf nodes probe for an empty location. When a leaf node's capacity is exceeded, a new node with extra capacity is created. 

Deletions follow the reverse procedure of insertions, a point query locates the key-value pair, which is then marked as available for future insertions. Child nodes are converted back to storage buckets once sufficient keys are deleted.

\section{Experimental Evaluation}
\label{sec:experiments}
All experiments are performed on a high performance computer with an Intel(R) Xeon(R) Gold 6448H CPU and 128 GB RAM. For the baselines, any training required for RL is performed on a system with an Intel(R) Xeon(R) Gold 6326 CPU with 128 GB of RAM, along with an A100 Nvidia GPU. 

We evaluate three variants of our \method\ structure, (1)~\textbf{\method}, the default variant, with hashing-based leaf nodes using a default threshold value set at 256 unless noted otherwise. The sensitivity of this threshold is analyzed in Section~\ref{sec:para_sec}; (2)~\textbf{\method-w/o-hash}, which disables the use of hashing-based leaf nodes; (3)~\textbf{\method-i-hash}, where both learned linear inner nodes and hashing-based inner nodes are used. This variant is used to evaluate the query performance gains and losses brought by using hashing-based nodes. The code is implemented in C++.

\subsection{Experimental Settings}
\textbf{Competitors.} We compare \method\ with Chameleon~\cite{guo2024chameleon}, LIPP~\cite{wu2021lipp}, SALI~\cite{ge2023sali}, ALEX~\cite{ding2020alex}, PGM~\cite{ferragina2020pgm}, and VEGA~\cite{li2025vega} to cover a variety of different approaches including conventional learned indexes, RL-based, hashing-based, and SIMD optimized structures. Hashing-based nodes and SIMD search nodes from these structures are  used as candidate node types for our index structure to demonstrate its generalizability.

CARMI~\cite{zhang2021carmi} and HIRE~\cite{zhang2026hire} are compared using their reported performance over ALEX. Both of them report smaller improvements than \method. CARMI reports 1.3 and 1.5 times better performance than ALEX on Books and Facebook respectively, while ours shows 4.5 and 7.0. HIRE on a read-heavy workload shows only 0.8 and 1.2 over ALEX for the same datasets.

\textbf{Datasets.} We use six datasets: OSM, Genome, Facebook, Books, Uniform,  and Lognormal, from two benchmark works~\cite{kipf2019sosd,wongkham2022updatable}, each containing 200 million keys. 

\begin{table}[t]
\caption{SAIL Parameters (default values in boldface)}
\begin{small}
    \label{tab:sail_paramters}
    \centering
\begin{tabular}{l l } 
\toprule
  Parameter & Values varied\\ \midrule
Type of nodes & Learned linear model, SIMD, Binary, Hash\\ \hline
Bucket size & 1, \textbf{3}, 6\\ \hline
Node size & 1, \textbf{1.65}, 2, 3\\ \hline
Threshold size & 0, \textbf{256}, 1024\\ 
\hline
 Key range&10, 100, 1000, 10 000, 100 000\\
 \bottomrule
\end{tabular}
\end{small}
\end{table}

\textbf{Parameters.}
There are several parameters that could potentially impact the performance of our proposed solution: (1)~\textbf{Type of nodes used}, the type of nodes available; (2)~\textbf{Bucket size}, the length of the buckets available in an inner node; (3)~\textbf{Node size}, the size of the inner nodes, expressed as a multiple of the size of the data modeled by that inner node. A default value of $1.65$ capacity is used for the benchmarking, following the best baseline index, Chameleon; and (4)~\textbf{Threshold size}, the maximum number of keys allowed in a hashing-based leaf node. Table~\ref{tab:sail_paramters} details the default parameters used as well as the values varied during the parameter analysis in Section~\ref{sec:para_sec}.

\textbf{Evaluation metrics.} We use all keys of a dataset to query the index structures after they are built. The keys are randomly shuffled to form the query set (same set for all indexes). For range queries, the start and end keys are randomly selected with a set key range. This is done for five different key ranges. 
We report: (1)~the \textbf{query time (ns)}, for point queries. (2)~the \textbf{index size (GB)}; (3)~the \textbf{build time (s)}, which is the time taken to build an index structure. (4)~the \textbf{range query time (ns/result)}, the total range query time divided by the number of keys within the queried range;  and (5)~the \textbf{insertion time~(ns)}.

\begin{figure}[t]
  \centering
  \includegraphics[width=0.9\linewidth]{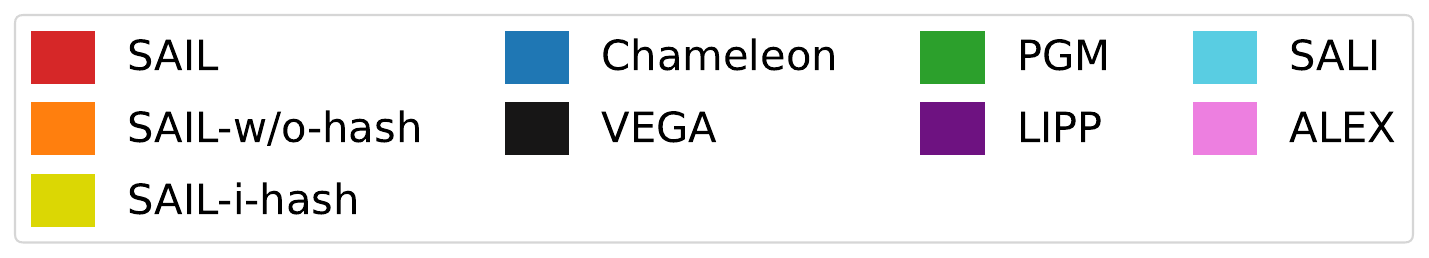}
  \includegraphics[width=\linewidth]{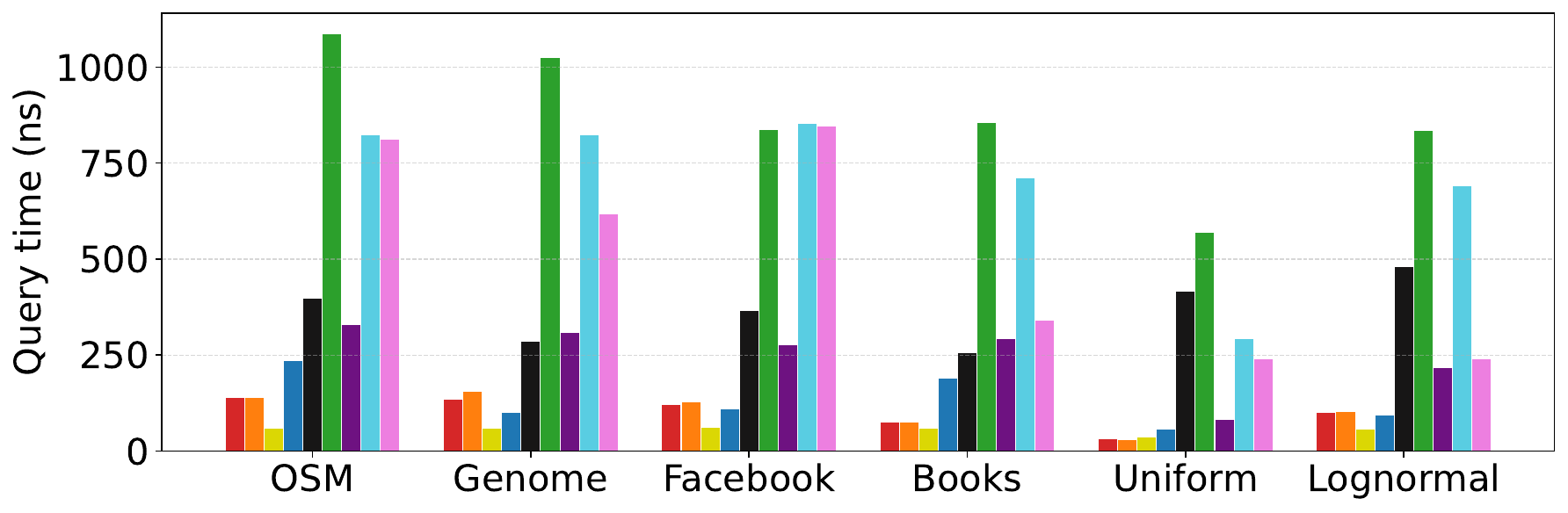}
  \vspace{-20pt}
  \caption{Point query time by dataset.}
  \label{fig:throughput}
  \vspace{-10pt}
\end{figure}

\subsection{Results on Read-only Workloads}
\label{sec:runtime}
For read-only workloads, each index is created using the full datasets, each with 
200 million data points.

\textbf{Point query time.} 
Fig.~\ref{fig:throughput} plots the point query times. \method\ and \method-w/o-hash are comparable or outperform the best baseline, Chameleon, with over 2 times better performance against all other baselines. Chameleon's strong query performance is due to its use of error bounded hashing at leaf nodes, which however severely hinders range query support. This trade-off is also observed by our \method-i-hash variant. Therefore, our evaluations will focus on the other two variants. \method\ marginally outperforms \method-w/o-hash except over the Uniform key-set. The Uniform key-set has a simple distribution, leading to identical single node indexes in both. The performance difference is due to random variation.

\begin{figure}[t]
  \centering
       \includegraphics[width=\linewidth]{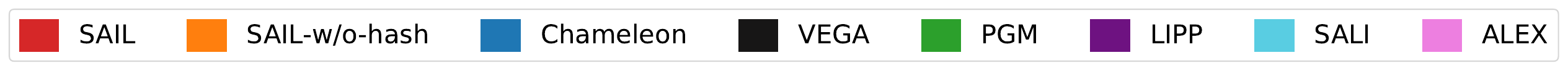}

    \includegraphics[width=\linewidth]{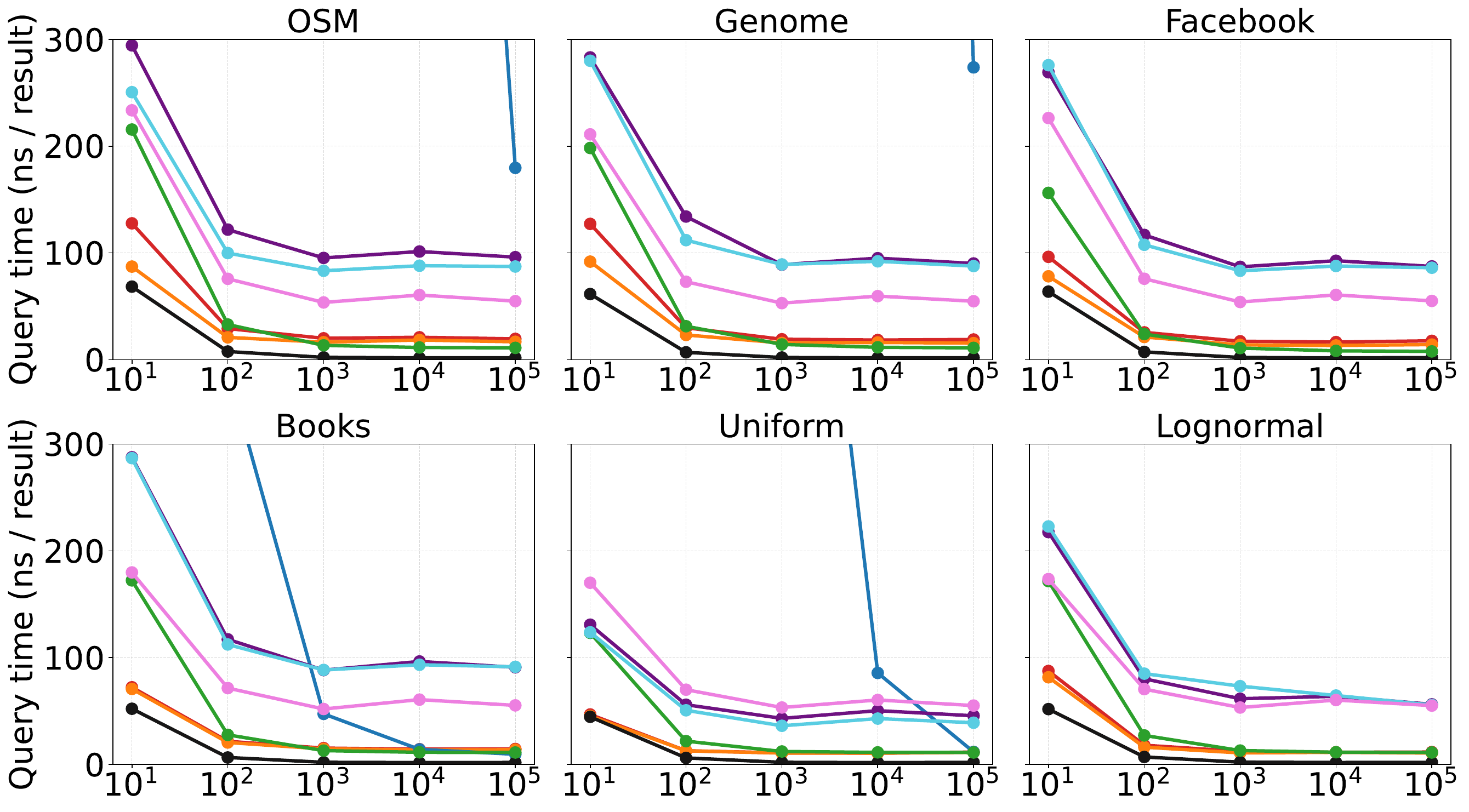}
  \caption{Range query time with different key ranges.}
 \label{fig:range_with_ranges}
 \vspace{-10pt}
\end{figure}

\textbf{Range query time.} Fig.~\ref{fig:range_with_ranges} shows range query times. Chameleon and \method-i-hash are cropped from Fig.~\ref{fig:range_with_ranges} because of their high range query times due to their hashing-based nodes. \method-i-hash is excluded from further evaluations, as \method\ already offers excellent point query performance without hindering range query efficiency.
All indexes become more efficient per result as the key range increases, since the main bottleneck is locating the leaf nodes. Afterwards, retrieving keys inside the same leaf node is faster. Chameleon's full node scan using linear search in these large hash nodes becomes better utilized as the range grows.

\method\ outperforms all baselines except VEGA and PGM. VEGA is the fastest due to its structure being optimized for read-only workloads, but it cannot support insertions. Although PGM supports insertions, it remains read-optimized, so its insertion performance suffers, which will be shown in Section~\ref{sec:insertion_sec}. Therefore, both indexes have significant limitations in real-world applications. For all baselines except Chameleon, efficiency stabilizes after the key range reaches 1,000. \method's hashing-based leaf nodes slightly hinders its range query efficiency at a key range of 10 (discussed further in Section~\ref{sec:para_sec}) for the same reason as Chameleon. This gap becomes negligible elsewhere with at most 9 ns difference against \method-w/o-hash on the OSM dataset. This shows that overall no baseline index performs consistently well on both point and range queries.

\begin{figure}[t]
  \centering
      \includegraphics[width=0.9\linewidth]{figures/legend2_v3.pdf}

    \includegraphics[width=\linewidth]{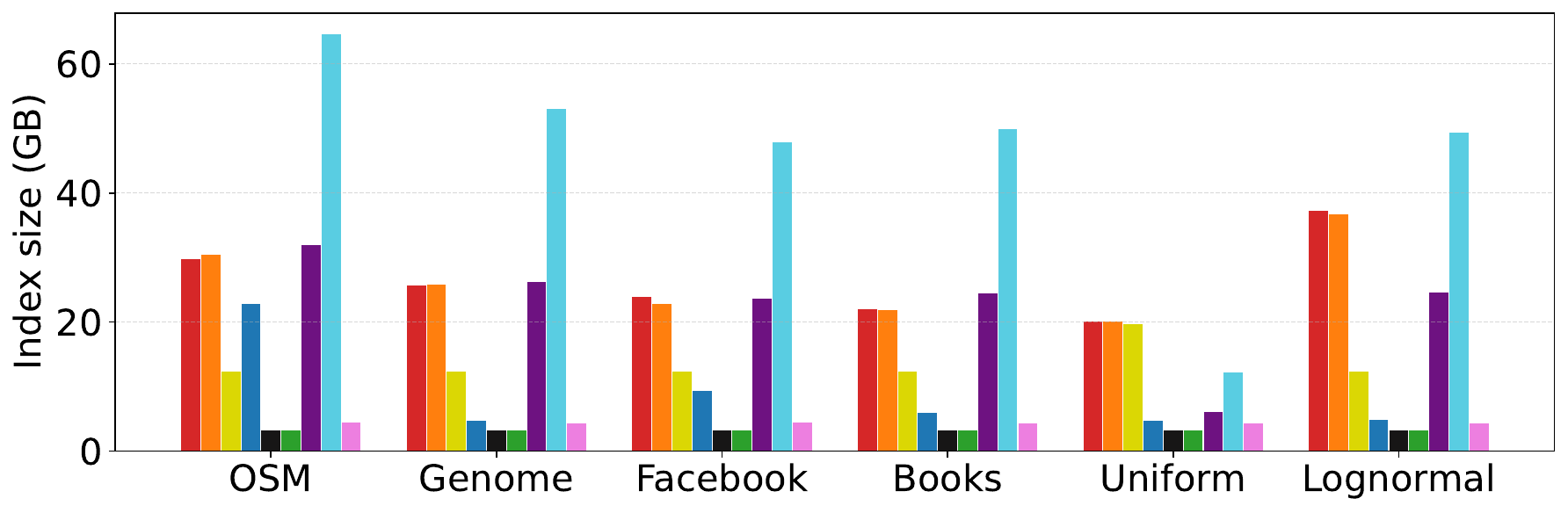}
    \vspace{-20pt} 
  \caption{Index size by datasets.}
 \label{fig:index_size}
 \vspace{-10pt}
\end{figure}

\begin{figure}[t]
  \centering

    \includegraphics[width=\linewidth]{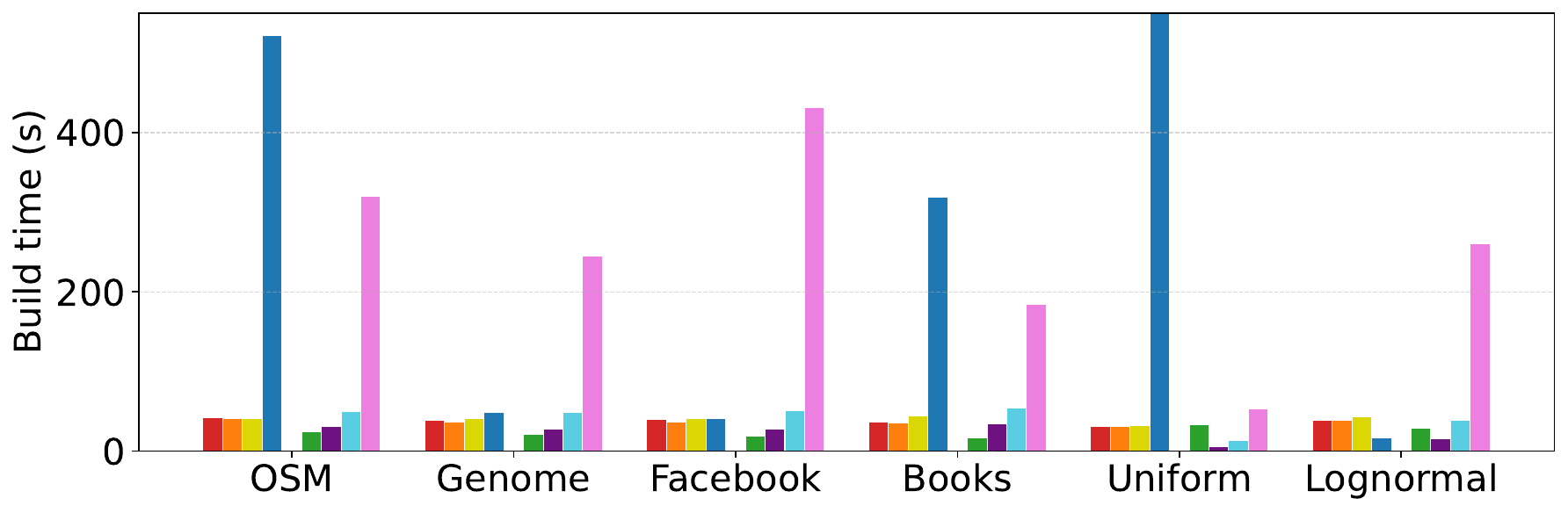}
    \vspace{-20pt} 
  \caption{Build time by datasets.}
 \label{fig:build_time}
\end{figure}

\textbf{Index size.} 
Fig.~\ref{fig:index_size} shows the index size. As expected, the size of \method\ is comparable to that of the LIPP index. On the datasets where this is deviated, such as Uniform and Lognormal, the performance gain of \method\ in query time is significant compared to the size increase, as shown in Fig.~\ref{fig:throughput}. The higher space consumption of \method\ in these scenarios can be attributed to the buckets introduced. The Lognormal dataset consumes the most memory space for \method, as the majority of its data are skewed. Even in this case, the index size of \method\ is only 1.5 times of that of LIPP while delivering 2.1 times faster point query processing (cf. Fig.~\ref{fig:throughput}).

\textbf{Build time.} 
Fig.~\ref{fig:build_time} shows the build time of the indexes. \method's  build times are consistent across all datasets and are comparable to the baselines. Chameleon and ALEX have high build times.  Chameleon's RL based construction requires up to 10,000 seconds even when measured on a GPU enabled device. ALEX's use of expensive cost models to recursively optimize the index contributes to its slow build time.

\begin{figure}[t]
  \centering
       \includegraphics[width=\linewidth]{figures/legend2.pdf}

    \includegraphics[width=\linewidth]{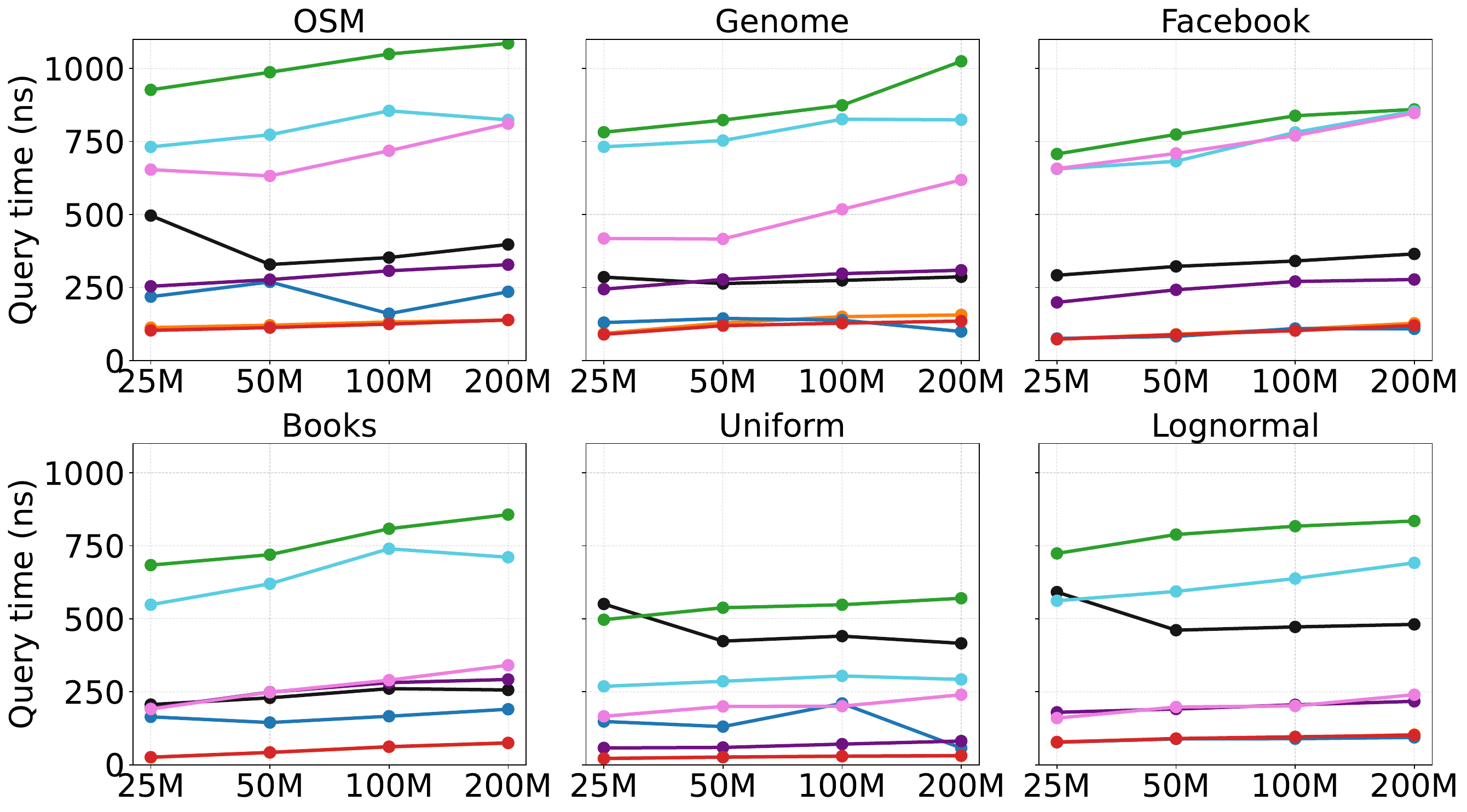}
  \caption{Point query time by data cardinality.}
 \label{fig:throughput_with_card}
 \vspace{-10pt}
\end{figure}
\begin{figure}[h]
  \centering

    \includegraphics[width=\linewidth]{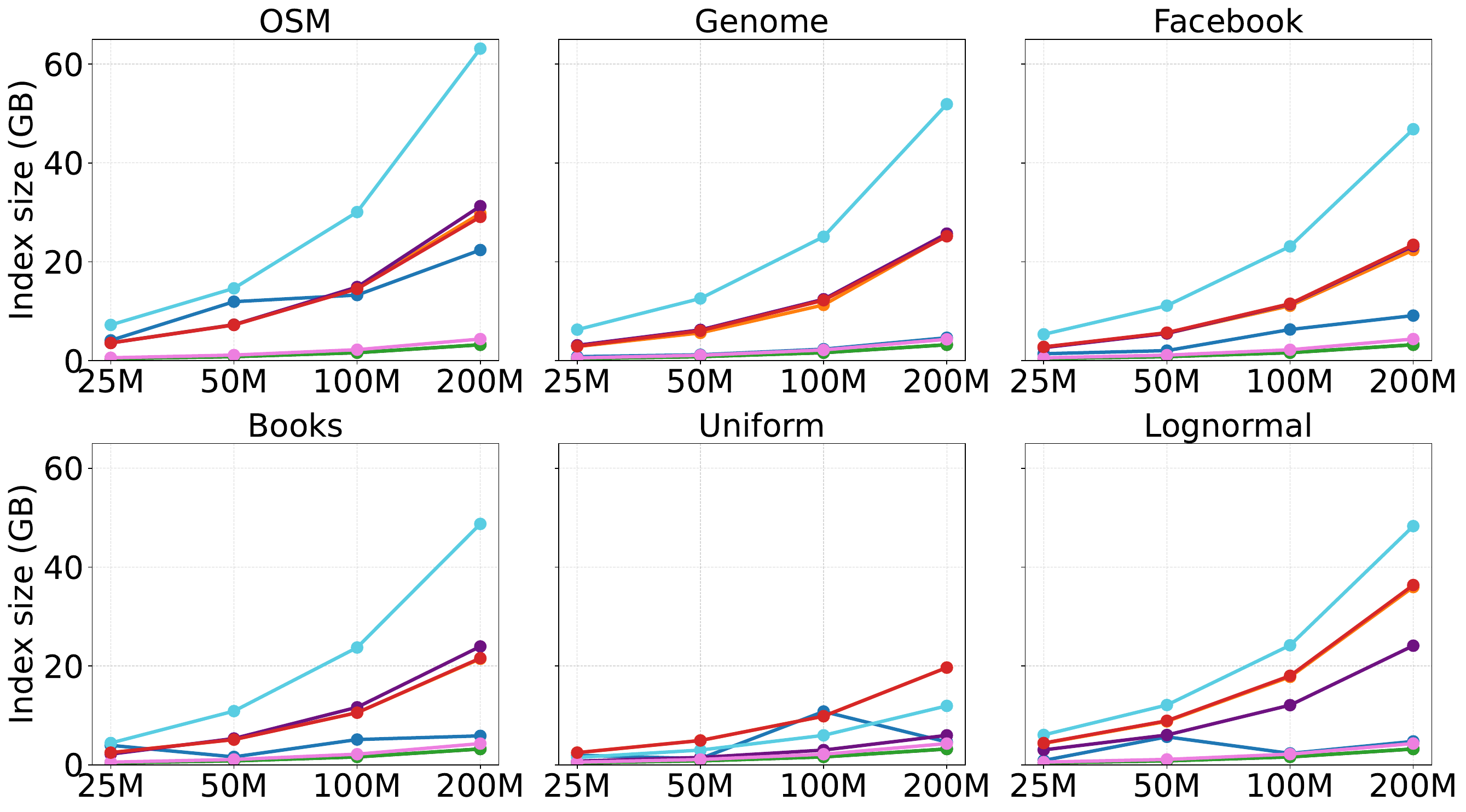}
  \caption{Index size by data cardinality.}
 \label{fig:index_size_with_card}
\end{figure}

\textbf{Imapct of dataset cardinality.} 
We evaluate the scalability of \method\ against different dataset cardinalates. Using the down sampling implementation from the SOSD benchmark~\cite{kipf2019sosd}, we generate three more datasets of sizes 25 million, 50 million and 100 million for each of the 200 million datasets. The dataset was down sampled by selecting keys at a fixed interval to obtain the required number of keys.

As seen in Fig.~\ref{fig:throughput_with_card}, for all index structures, as the dataset cardinality increases, the query performance tends to degrade, except for Chameleon due to the non-deterministic nature of the index structure. The degradation of \method\, when taken as a percentage with the 25 million and 200 million datasets, is comparable to most of the baseline index structures. VEGA  is the most stable among all baselines. However, for all of the dataset sizes, our \method\ index's query performance remains the best by a large margin except for Chameleon, for which the reasons were discussed earlier on point query performance. 

Fig.~\ref{fig:index_size_with_card} shows the index size as the dataset cardinality varies. As expected, the index size increases as the dataset size increases. The indexes that have the ability to hold keys in their inner nodes, i.e., \method, LIPP, and SALI have the highest index space consumption, as they share similar architectures. 

\begin{figure}[t]
  \centering
       \includegraphics[width=\linewidth]{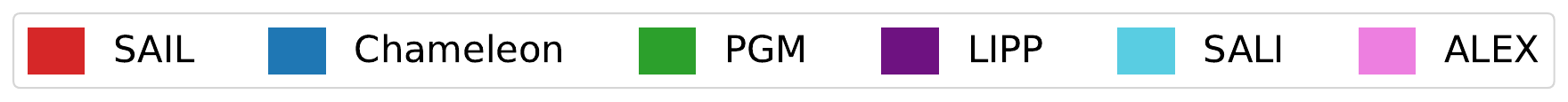}

    \includegraphics[width=\linewidth]{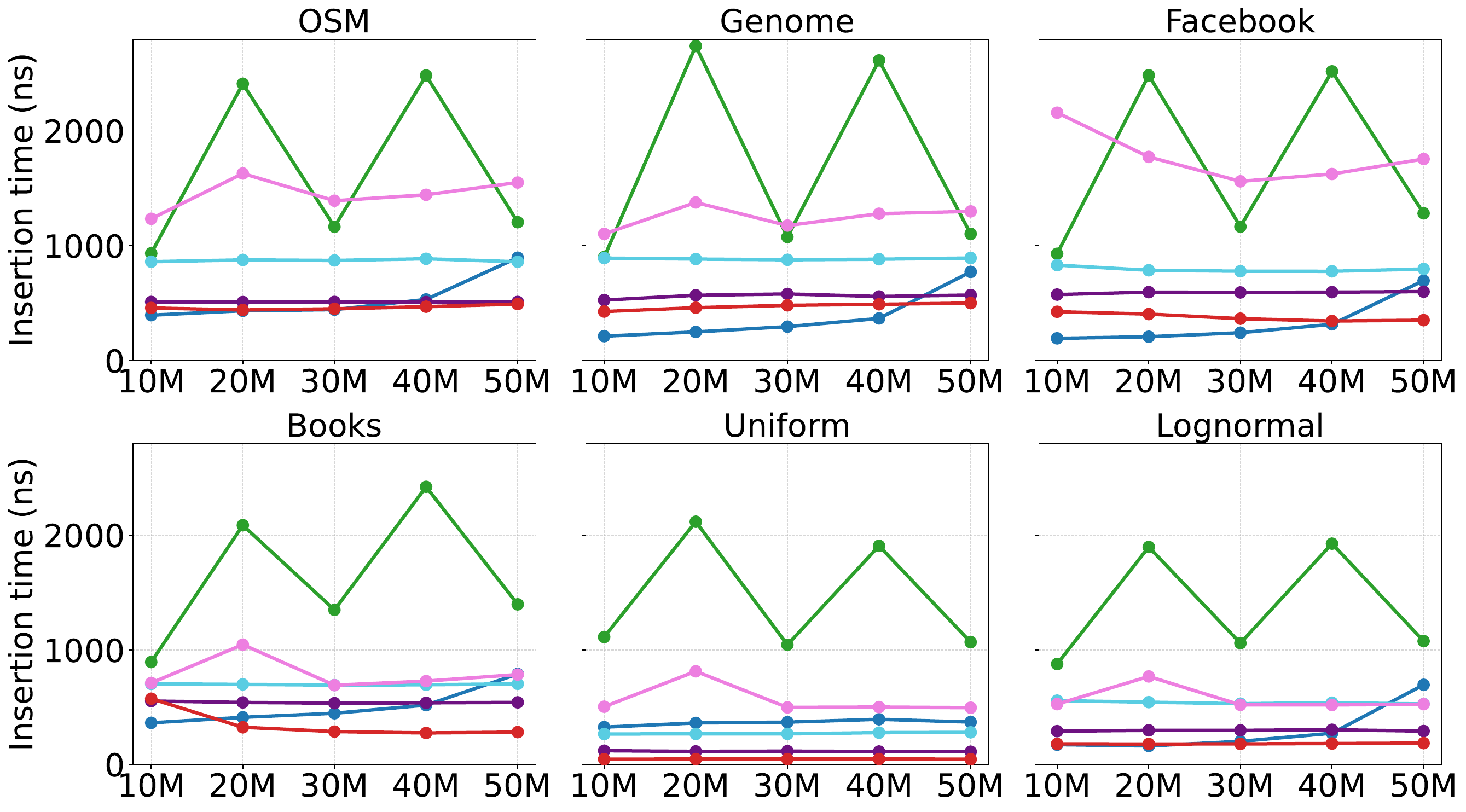}
  \caption{Insertion time by insert batches.}
 \label{fig:insert_time}
 \vspace{-10pt}
\end{figure}

\begin{figure}[t]
  \centering

    \includegraphics[width=\linewidth]{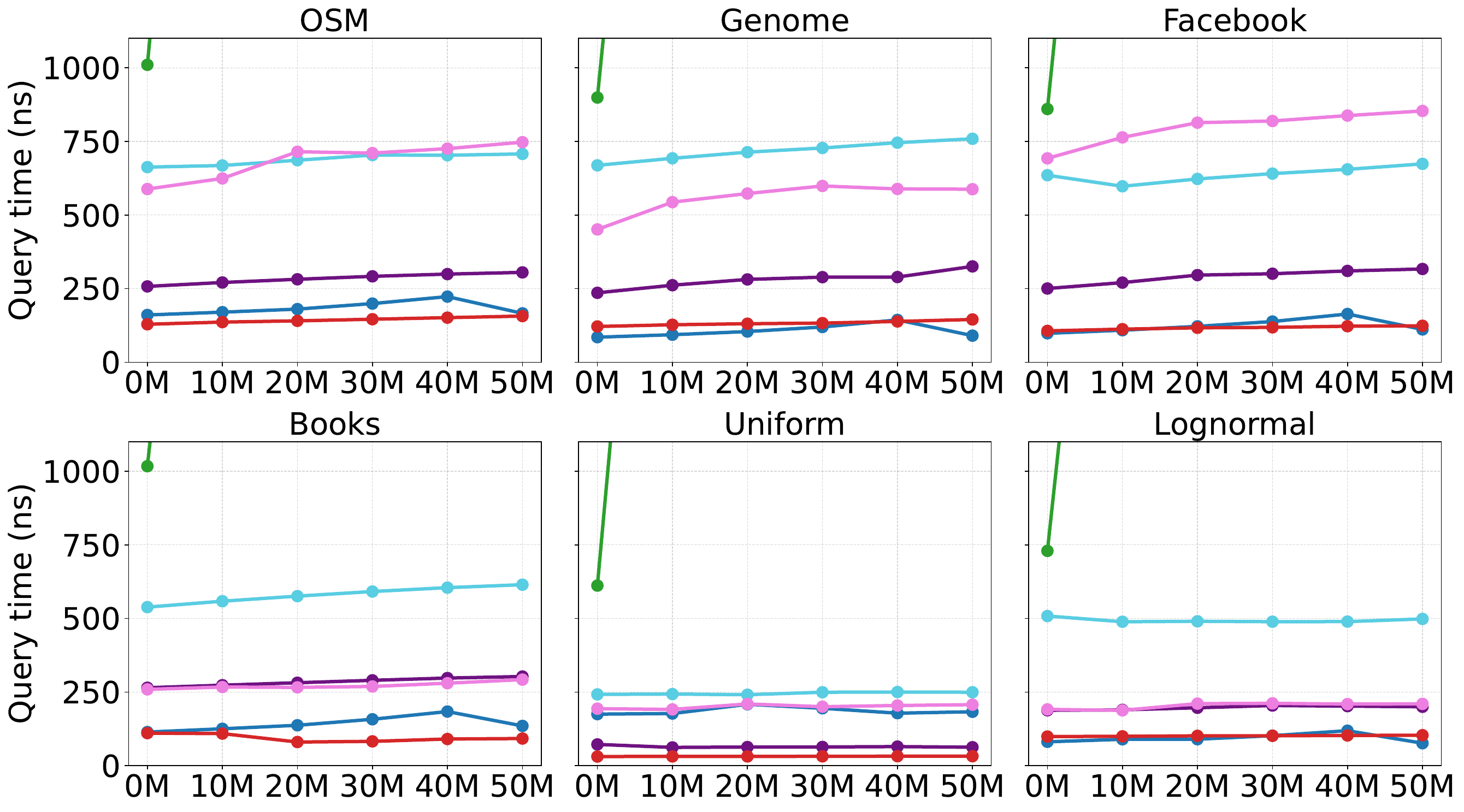}
  \caption{Query time after insertions in batches.}
 \label{fig:insert_query_time}
 \vspace{-10pt}
\end{figure}

\subsection{Results on Read-Write Workloads}
\label{sec:insertion_sec}
For the Read-Write workloads, key-sets of 100 million data points are used to bulk load the indexes. From the remaining 100 million data, five batches of 10 million data points are taken randomly such that no keys are repeated. The performance of the index is measured after inserting each batch.  VEGA  is removed from this experiment due to its inability to support insertions. \method-w/o-hash is also removed from this section onward to reduce redundancy, as  \method\  has shown better performance already.

\textbf{Insertion time.} 
Fig.~\ref{fig:insert_time} shows the insertion time in nanoseconds for the  six key-sets. SAIL performance is comparable or better than the best performing baseline, Chameleon, without requiring pretrained RL agents. PGM  shows the worst performance, as it was not designed for insertions. The fluctuating patten it displays can be attributed to the merging it performs to add the new keys. The benefit of this merging is passed on to the next insertion batch, allowing it to perform the insertions relatively faster in that batch. 
\method, LIPP, and SALI show stable insertion times across the batches, while Chameleon shows an increasing pattern.

\textbf{Point query time.} 
Fig.~\ref{fig:insert_query_time} shows the zoomed in plots for the query time after each of the insert batches are added. This is done due to the PGM index displaying substantially worse performance. \method\ shows consistent performance throughout the five insert batches, with only a gradual increase. Most baselines tend to show a steeper decline in query performance. Chameleon performance decreases in a steady pace till the final insert batch, after which a rebuild operation occurs. Doing so aids its query performance.

\begin{figure}[t]
  \centering
      \includegraphics[width=0.7\linewidth]{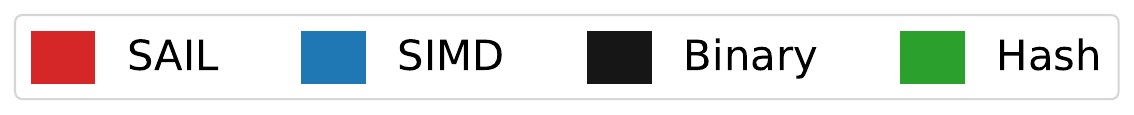}
    \includegraphics[width=\linewidth]{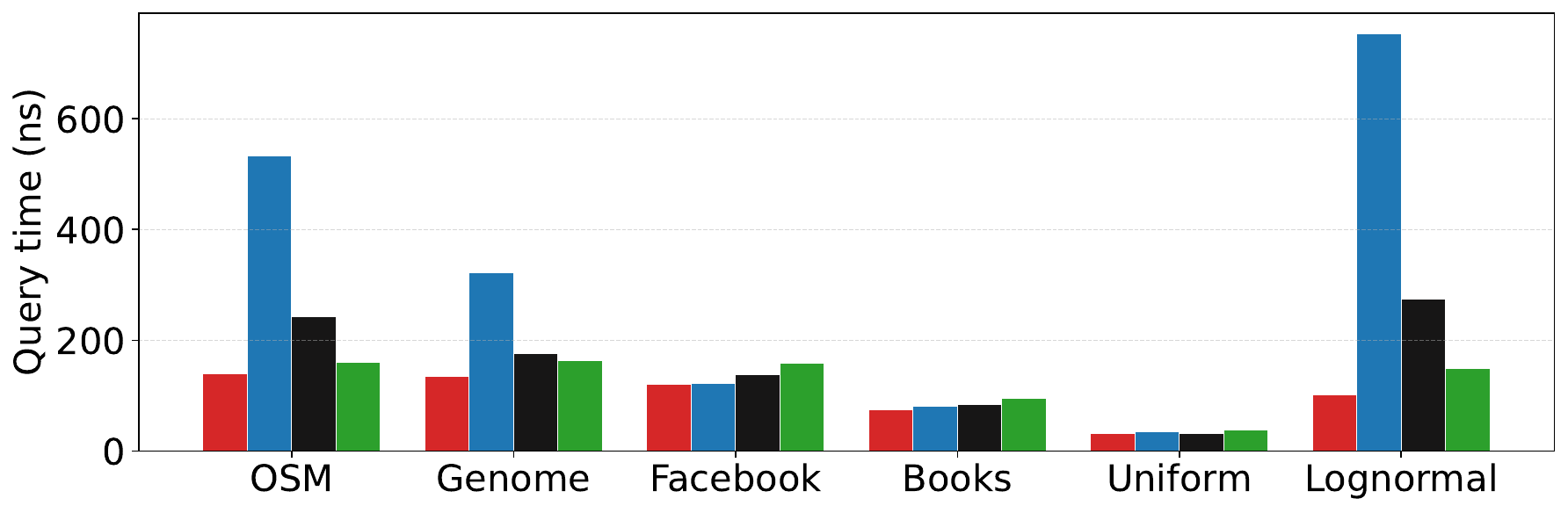}
  \caption{Point query time with \method\ using single types of nodes.}
 \label{fig:throughput_sail_nodes}
\vspace{-5pt}
\end{figure}

\begin{figure}[t]
  \centering
      \includegraphics[width=\linewidth]{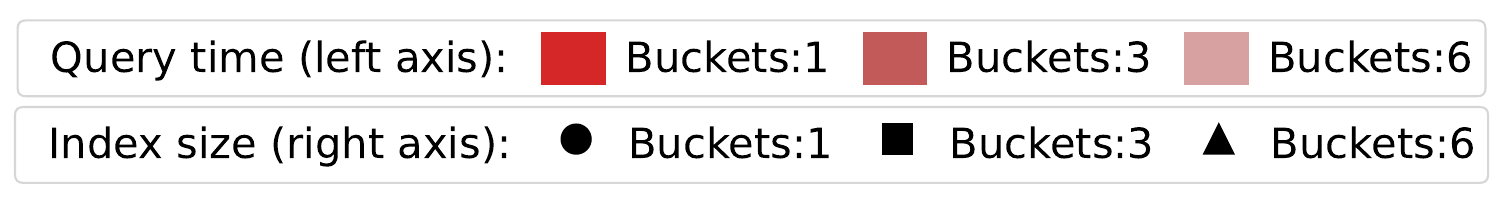}
    \includegraphics[width=\linewidth]{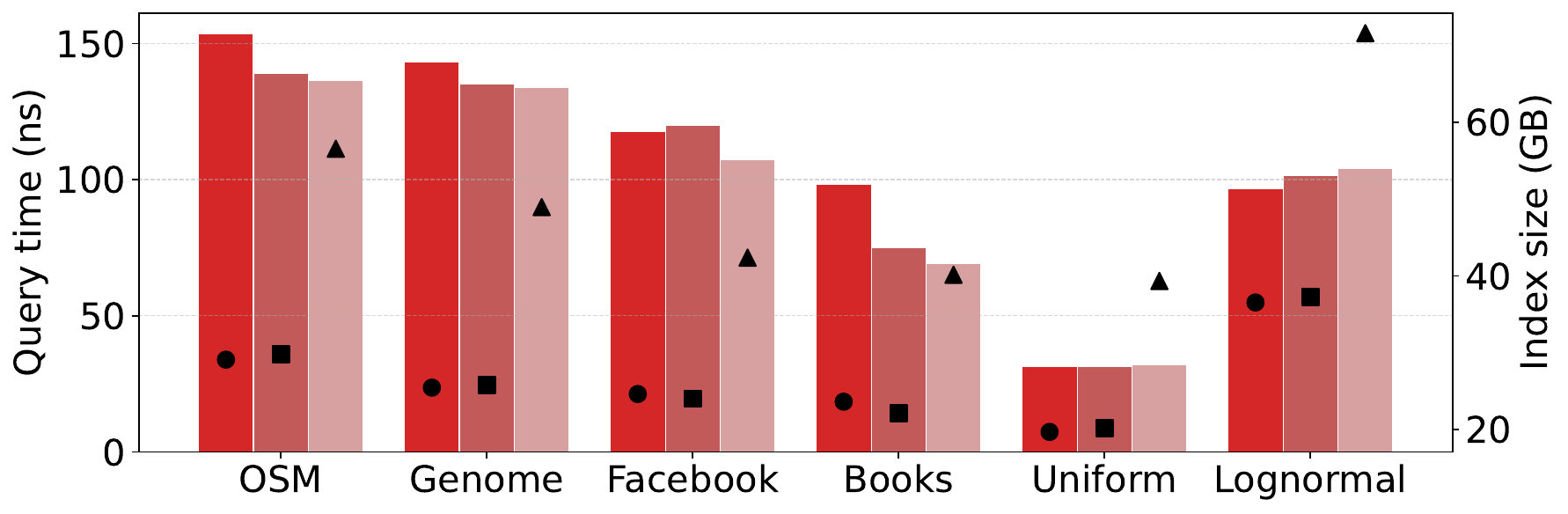}
  \caption{Point query time and index size by bucket sizes.}
 \label{fig:sail_buckets}
\end{figure}

\begin{figure}[t]
  \centering
      \includegraphics[width=\linewidth]{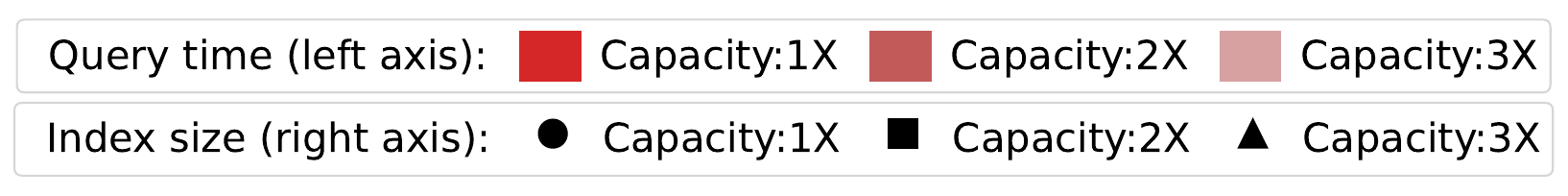}
    \includegraphics[width=\linewidth]{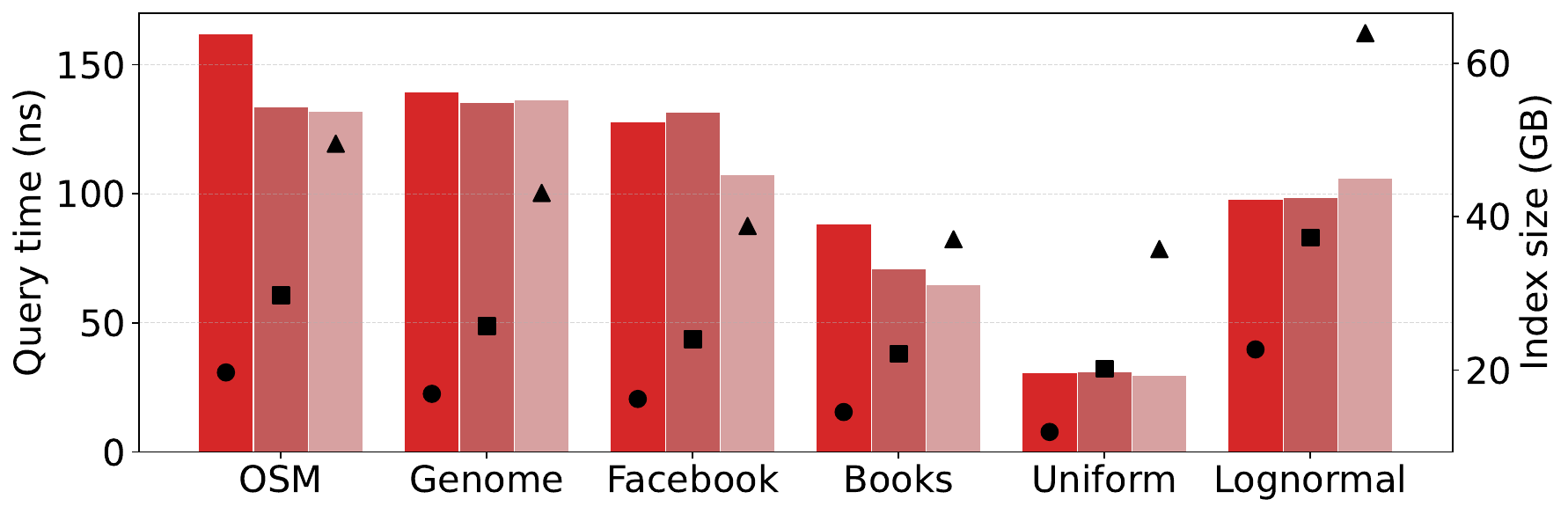}
  \caption{Point query time and index size by node capacity.}
 \label{fig:sail_capacity}
\end{figure}

\begin{figure}[t]
  \centering
      \includegraphics[width=\linewidth]{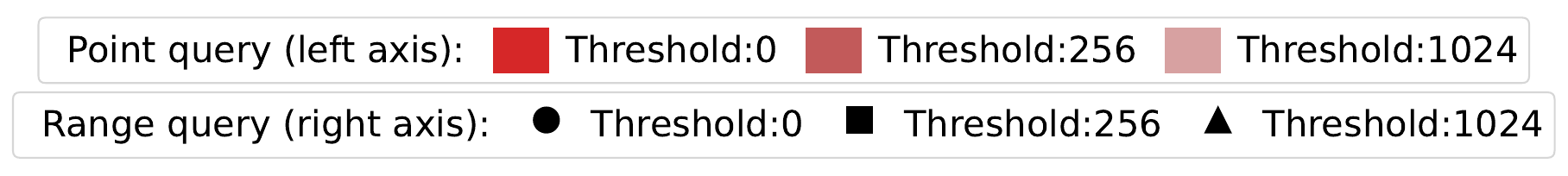}

    \includegraphics[width=\linewidth]{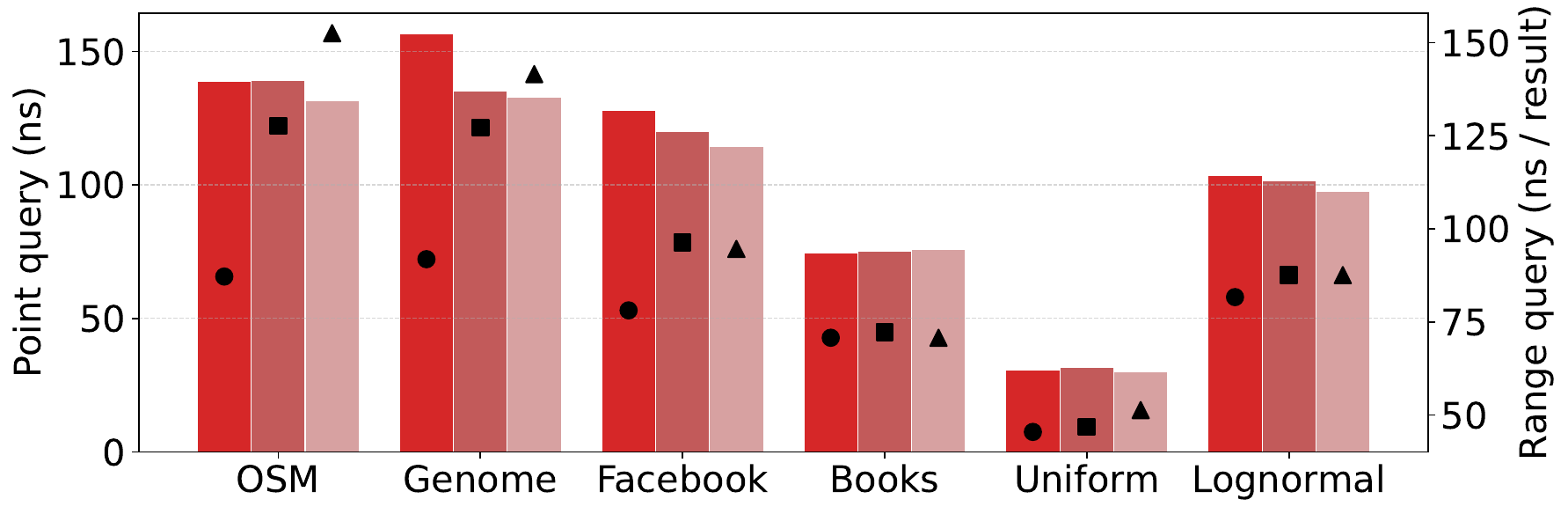}
    \vspace{-10pt}
  \caption{Point query and range query time by hashing node sizes.}
 \label{fig:sail_threshold_point}
\end{figure}

\subsection{Parameter Analysis}
\label{sec:para_sec}
We vary the parameters of our proposed \method\  index structure to analyze their impact on the query time and index size. We vary each parameter individually while keeping all others to their default values as given in Table~\ref{tab:sail_paramters}. 

\textbf{Node type.}
We disable all other node types except one to study the impact of the node types. Due to the large size of key-sets, the root node was fixed with a learned linear inner node. Using just  learned linear nodes for the entire index is omitted, as it costs prohibitive memory space consumption. Fig.~\ref{fig:throughput_sail_nodes}  shows that using all available node types delivers the lowest query time. For \method, the availability of multiple node structures could lead to query performance degradation, if the cost model failed to correctly identify the most appropriate node structure. As such, the results here also demonstrate the effectiveness of the cost models.

\textbf{Bucket size.}
The bucket size of the inner node type can be increased to hold more overflowing keys at specific locations. However, doing so would significantly increase the memory consumption. As shown by Fig.~\ref{fig:sail_buckets}, bucket size 3 offers the best performance to memory trade-off while for size 6, memory consumption is greater for a marginal query performance improvement. On Lognormal data, bucket size has little impact as the majority of the keys are concentrated in a small region. Bucket sizes 1 and 3 have similar index sizes because the memory saved in the inner nodes for bucket size 1 is offset by the additional leaf nodes needed for the overflowing keys.

\textbf{Node capacity.}
\label{sec:node_cap}
We vary the capacity of the inner node type by 1 times, 2 times and 3 times the number of data points in the partition. Increasing the capacity allows the indexing function to place more keys in non-overflowing locations, at the cost of extra memory. As shown in Fig.~\ref{fig:sail_capacity}, this increase in the node capacity correlates to better query performance in most cases. Increasing node capacity does not aid on the Lognormal key-set for the same reason as in the higher bucket size scenario. A gradual increase in the index size can be seen when higher capacity nodes are used. Capacity of 2 times the key-set delivers the most performance gain for a minor memory increase. 

\textbf{Threshold size.}
To measure the impact of the hashing-based leaf node on point and range query performance, we adjust its maximum size. Fig.~\ref{fig:sail_threshold_point} shows that increasing the size of  hashing-based leaf nodes improves the point query performance marginally, but at the cost of its range query performance. This is also shown by Fig.~\ref{fig:sail_threshold_point} for key range 10. Therefore, a threshold size of 256 gives the best balance between point and range query performance. 

\section{Conclusion}
We addressed the issue of inefficient indexing of key-sets due to their localized distribution variations. We proposed \method, a self-adaptive index structure, using different types of nodes to better capture the distributions of specific key partitions. \method\ recursively partitions a key-set and assigns appropriate node types to these partitions. We evaluated the index on four real-world key-sets and two synthetic key-sets. \method\ reported  approximately 2x improvement over order-preserving learned indexes in query time with a similar memory and build time. It also demonstrated on par point query performance with hashing-based learned index structures. 

Overall, our study presents empirical evidence for the adoption of hybrid index structures with 
multiple node types, showing that such structures lead to better query efficiency as key-sets grow in size and distribution complexity. 
For future work, we plan to extend \method\ to a workload-aware structure by incorporating workload information into the cost models.

\section*{Acknowledgment}
Claude was used to assist in the implementation and improving the readability of the text. The authors take full responsibility for the final content of the paper. 

\bibliographystyle{ieeetr}
\bibliography{sample-base}

\vspace{12pt}

\end{document}